\documentclass[aps,pra,reprint,superscriptaddress,nofootinbib]{revtex4-2}

\usepackage[T1]{fontenc}
\usepackage[utf8]{inputenc}
\usepackage{amsmath,amssymb,amsthm,mathtools}
\usepackage{braket}
\usepackage{bm}
\usepackage{hyperref}
\usepackage{cleveref}
\usepackage{enumitem}
\usepackage{xcolor}
\usepackage{graphicx}
\usepackage{tikz}
\usepackage{quantikz}
\usepackage{orcidlink}
\usepackage{appendix}

\graphicspath{{./images/},{./imagesAppendix/}}

\definecolor{THc}{rgb}{0.9,0.3,0.2}

\definecolor{ACc}{rgb}{0.2,0.3,0.9}

\hypersetup{colorlinks=true,linkcolor=blue,citecolor=blue,urlcolor=blue}

\newcommand{\synd}{\mathbf{s}}

\newcommand{\pL}{p_{\mathrm{L}}}
\newcommand{\pth}{p_{\mathrm{th}}}
\newcommand{\pml}{p_{\mathrm{ML}}}
\newcommand{\cpred}{C}

\newcommand{\err}{\mathbf{e}}

\newcommand{\idg}[1]{{\bfseries #1)}}

\newcommand{\subfigimg}[3][,]{%
	\setbox1=\hbox{\includegraphics[#1]{#3}}%
	\leavevmode\rlap{\usebox1}%
	\rlap{\hspace*{2pt}\raisebox{\dimexpr\ht1-0.5\baselineskip}{{\bfseries \large\textsf{#2}}}}%
	\phantom{\usebox1}%
}

\newcommand{\SM}{SM}

\begin{document}

\title{Machine-learned syndrome post-selection for reliable quantum error correction}

\author{Tobias Haug}
\email{tobias.haug@tii.ae}
\affiliation{Quantum Research Center, Technology Innovation Institute, Abu Dhabi, UAE}

\author{Askery Canabarro}
\affiliation{Quantum Research Center, Technology Innovation Institute, Abu Dhabi, UAE}
\affiliation{Campus Arapiraca, Federal University of Alagoas, Arapiraca-AL, Brazil}

\author{Leandro Aolita}
\affiliation{Quantum Research Center, Technology Innovation Institute, Abu Dhabi, UAE}

\date{\today}

\begin{abstract}
Quantum error correction can be enhanced
by post-selecting out
runs that are likely to produce a logical failure, but the most accurate measures for that 
require costly decoder-level information. We introduce a practical, decoder-agnostic post-selection method that learns directly from syndrome data. The method trains a supervised classifier to distinguish between syndromes from low- and high-noise regimes, and then uses the classifier's output as an abort score for new runs, without requiring logical-error labels, correction operators, or code-specific likelihood calculations. We validate
the approach in three complementary settings: circuit-level simulations of the Gross bivariate-bicycle code, code-capacity simulations of the surface code, and experimental logical magic-state distillation data from the QuEra neutral-atom processor. In the Gross and surface codes, learned syndrome post-selection reduces the conditional logical error rate at a fixed acceptance rate, with performance comparable to syndrome-weight filtering. For the surface code, the learned classifier reveals a post-selection transition distinct from the conventional decoding threshold. In the experimental data, the machine-learning score outperforms syndrome-weight post-selection and, when combined with logical-gap filtering, improves the output fidelity beyond using the logical gap alone. These results show that syndrome-only learning provides a scalable and hardware-compatible route to improving the reliability of quantum error correction.

\end{abstract}

\maketitle

\let\oldaddcontentsline\addcontentsline%
\renewcommand{\addcontentsline}[3]{}%

Quantum error correction (QEC) protects fragile quantum information by encoding logical information into a large number of physical qubits~\cite{terhal2015quantum}.  Stabilizer measurements diagnose physical noise through syndrome data, which are then processed by a classical decoder to choose a correction while preserving the encoded logical state.  In practice, however, some error patterns remain ambiguous and can evade the decoder, producing logical failures that spoil the computation.  This limitation is particularly relevant for near-term and early fault-tolerant devices, where the achievable logical performance is constrained by the number of available qubits, the physical error rate, the fidelity of syndrome extraction, and the speed and accuracy of classical decoding.  Methods that improve the conditional reliability of QEC without changing the quantum hardware are therefore valuable.

Post-selection provides one such method.  Rather than using every experimental shot, a post-selection rule aborts runs whose syndrome history suggests a high probability of logical failure~\cite{chen2022calibrated}.  If the abort rule is sufficiently predictive, the conditional logical error rate of the accepted ensemble can be much lower than the unconditional logical error rate~\cite{knill2005quantum,aliferis2007accuracy}.  This trade-off is especially useful when high-fidelity outputs are more valuable than deterministic throughput, as in logical magic-state preparation, magic-state distillation, magic-state cultivation, and repeat-until-success gadgets~\cite{bravyi2005universal,bravyi2012magic,bombin2024fault,gidney2024magic,akahoshi2024partially,staples2026scalable}.

A powerful route to post-selection is to augment decoding with soft information: instead of outputting only a correction, the decoder also estimates the reliability of that correction.  A prominent example is the \emph{logical gap}~\cite{bombin2024fault,meister2024efficient,chen2025scalable,smith2024mitigating,gidney2025yoked,sales2025experimental,lee2026efficient}, defined as the difference between the score of the most likely logical class and that of the second most likely logical class after decoding.  Intuitively, a small gap means that the decoder is uncertain, and such shots are promising candidates for rejection.  
However, logical-gap post-selection requires decoder-level likelihood information.  For structured decoders and codes, efficient approximations to such soft information are available~\cite{meister2024efficient}; for general qLDPC codes, experimental decoding graphs, or maximum-likelihood logical-class comparisons, obtaining reliable likelihood information can remain computationally demanding~\cite{hsieh2011np,iyer2015hardness}.  Thus, the most informative confidence metric may be precisely the one that is least convenient to obtain.  Alternatively, one can use decoder-free metrics such as aborting when the number of flipped syndrome bits exceeds a cutoff~\cite{li2015magic,singh2022high,bombin2024fault,english2025thresholds}, but such scores discard geometric, temporal, and correlation information contained in the full syndrome record.

Machine learning has also become a powerful tool for QEC, primarily through learned decoders that map syndrome histories to correction operators, logical classes, or decoder confidence estimates~\cite{torlai2017neural,krastanov2017deep,baireuther2019neural,meinerz2022scalable,bauch2024learning,lange2025datadriven,maan2025machine,gicev2025fully,dentelski2026neuralnetworkdecoderconfidence,gu2026scalable}.  Such methods can improve decoding accuracy on realistic noise and experimental data, but they typically require training data labeled by corrections, logical outcomes, or decoder-derived targets, and their architecture is often tied to a particular code family or noise model.  Here we take a different route: rather than learning to decode, we learn when not to decode.

\begin{figure}[t!]
    \centering
        \subfigimg[width=0.49\textwidth]{}{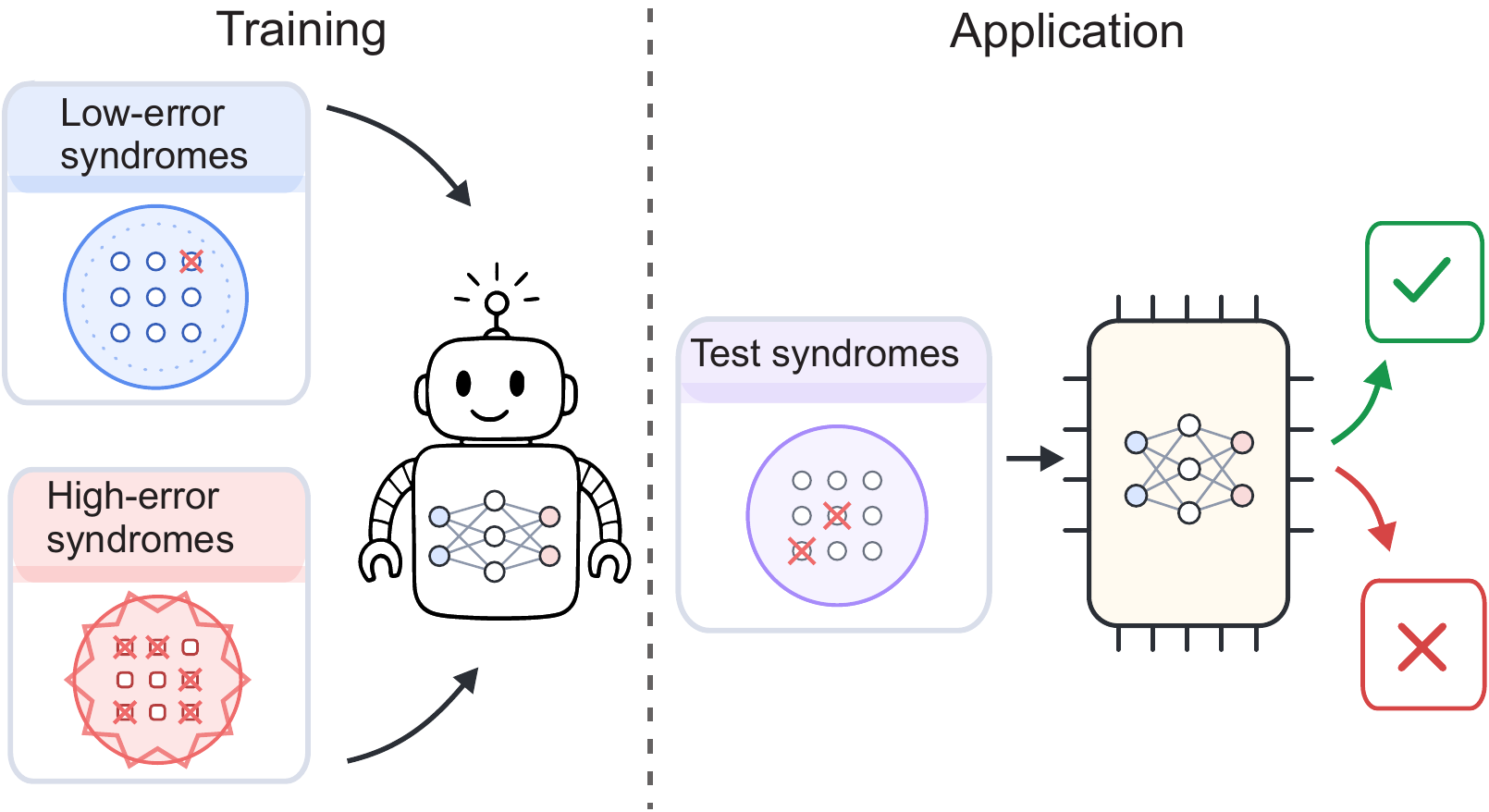}
    \caption{{\bf %
    Machine-learning post-selection via syndrome data}. The supervised machine-learning model learns from an %
    ensemble of syndromes with high- and low-error events. After training, the model is applied on new syndromes to decide which runs to actively correct errors on and which ones to discard. }
    \label{fig:sketch}
\end{figure}

We introduce a machine-learning method that learns directly from syndrome distributions to post-select.  The training data are simple to acquire (see Fig.~\ref{fig:sketch}): one generates two syndrome ensembles, one in a low-noise regime and one in a high-noise regime.  A supervised classifier is trained only to distinguish these two ensembles.  For a new syndrome $\synd$, the classifier output is interpreted as a high-noise score, and the post-selection rule aborts when this score exceeds a cutoff.  Crucially, the labels used for training are not logical-success or logical-failure labels, nor correction operators or logical-gap values, as in many machine-learning approaches~\cite{torlai2017neural,krastanov2017deep,baireuther2019neural,meinerz2022scalable,bauch2024learning,lange2025datadriven,maan2025machine,gicev2025fully,dentelski2026neuralnetworkdecoderconfidence,seip2026machine,gu2026scalable}. Instead, our labels only specify from which noise ensemble the syndrome was sampled.  The method is therefore decoder-agnostic and can be trained from simulation, from calibration data at deliberately varied noise rates, or from a mixture of both.

We benchmark this syndrome-only post-selection strategy in three settings.  First, we train on simulated syndrome data for a neutral-atom logical magic-state distillation experiment~\cite{sales2025experimental} and apply the resulting classifier to actual experimental data.  In this case, our machine-learning post-selection method outperforms syndrome-weight filtering. Further, by combining our method with logical-gap post-selection, we achieve higher output fidelity than the original logical-gap filtering method of Ref.~\cite{sales2025experimental}.
Second, we apply it to circuit-level simulations of the $[\![144,12,12]\!]$ Gross bivariate-bicycle code, demonstrating that the method applies beyond standard surface-code decoding graphs.  As a third application, we study the surface code under a code-capacity noise model and find that the learned score exhibits a crossing at a post-selection threshold $\pml$ distinct from the usual logical-error threshold $\pth$~\cite{english2025thresholds}. Our results identify syndrome-distribution learning as a practical and modular pre-decoding filter for improving the conditional reliability of QEC.

\section{Preliminaries}
\label{sec:preliminaries}

\subsection{Syndromes and decoding}

QEC codes are characterized by parity check matrix $H \in \mathbb{F}_2^{m\times n}$, in which syndrome measurements produce the $m$-bit syndrome $\synd \in \{0,1\}^{m}$ given by %
\begin{equation}
    \synd=H \err 
\end{equation}
which is the result of $n$ possible errors in error configuration $\err\in \{0,1\}^{n} $ that affected the code~\cite{spencer2026quantum}. %
Then, a most-likely-error decoder tries to find the most probable physical error $\err^\ast$ that matches the syndrome.  
From this, the correction operator is determined and applied, where the decoder succeeds when the correction returns the state to its original form without inducing a logical error. For example, a CSS stabilizer code $[\![n,k,d]\!]$ with code distance $d$ can protect $k$ logical qubits against up to $\lfloor d-1\rfloor/2$ errors.

However, the decoding problem is notoriously hard~\cite{hsieh2011np,iyer2015hardness}. Further, the most probable error may not be the actual error that has occurred. This can lead to a corruption of the logical state, i.e. a logical error. The logical error probability (after decoding) at physical error rate $p$ is denoted by $\pL$. 

Undetected logical errors can corrupt the underlying computation. As such, when  one has strong indication that decoding is unlikely to succeed, it is often better to simply discard the computation, and restart with a fresh run. 
Such post-selection strategies are also essential in magic state distillation protocols, where one distills resource states from noisy states~\cite{bravyi2005universal,bravyi2012magic}, or repeat-until-success schemes~\cite{akahoshi2024partially}.
In such post-selection schemes, only a subset $\tilde{N}_\text{shots}$ of the total number of shots $N_\text{shots}$ is used, while the rest is discarded even before decoding. 
We denote by
\begin{equation}
    R=\frac{\tilde{N}_\text{shots}}{N_\text{shots}}
    \label{eq:postselectionrate}
\end{equation} 
the post-selection acceptance rate, i.e. %
the probability that our post-selection rule accepts shots, and by %
$%
\tilde{p}_L$ %
the logical error rate after post-selection and decoding.

\subsection{Post-selection and the logical gap}

A post-selection rule is a scalar score $g(\synd)$ together with a cutoff $\tau$ that is used to discard QEC runs that are deemed unreliable.  The rule accepts a shot when $g(\synd) < \tau$ and aborts otherwise.  
Several schemes have been proposed so far. 
A simple decoder-free score is the syndrome-weight~\cite{english2025thresholds}
\begin{equation}
    w(\synd)=\sum_{i=1}^{m} s_i,
\end{equation}
which counts the number of nontrivial detection events, i.e. of %
flipped syndrome bits.  Syndrome-weight post-selection is computationally cheap and can be surprisingly effective, but it discards detailed information about the geometry, correlations, and temporal structure of the syndrome record.

A more informed score is the logical gap~\cite{bombin2024fault,meister2024efficient,chen2025scalable,smith2024mitigating,gidney2025yoked,sales2025experimental,lee2026efficient}.  Let $L_1(\synd)$ and $L_2(\synd)$ be the largest and second-largest probability of physical errors that match syndrome $\synd$ and belong to different logical cosets.%
Then, the logical gap is
\begin{equation}
    \Delta(\synd)=\log L_1(\synd)-\log L_2(\synd).
    \label{eq:gap}
\end{equation}
where larger $\Delta$ indicates a more confident decoding decision.  
Logical-gap post-selection is attractive because it directly measures the ambiguity relevant to logical failure.  Its drawback is that estimating $L_1$ and $L_2$ can be a hard decoding problem.

\section{Machine-learning post-selection}
\label{sec:method}

We now introduce our machine-learning approach to syndrome-based post-selection. The central idea is to train a binary classifier to distinguish syndromes generated in low- and high-error-rate regimes. The resulting classifier score is then used as a decoder-independent measure of whether a given syndrome resembles the high-noise ensemble and should therefore be rejected.

\subsection{Supervised learning from syndrome ensembles}

Let $\mathbf{s} \in \{0,1\}^{m}$ denote a syndrome record, or a
vectorized syndrome history when several rounds of syndrome extraction
are considered. We formulate syndrome post-selection as a supervised
binary-classification problem in which the target label identifies the
noise ensemble from which each syndrome was sampled:

\begin{equation}
y =
\begin{cases}
0, & \text{low-error-rate ensemble},\\
1, & \text{high-error-rate ensemble}.
\end{cases}
\label{eq:ml_labels}
\end{equation}
Importantly, these labels do not indicate whether decoding succeeds or
whether a logical error occurs. They only specify the physical-error-rate
regime used to generate each syndrome.

A classifier $C(\mathbf{s})$ is trained to distinguish the two syndrome
ensembles. 
The classifier predicts the probability $\text{Pr}(y=1\mid\mathbf{s})$ that a given syndrome $\mathbf{s}$ belongs to class $y=1$, which corresponds to the high-error-rate dataset
\begin{equation}
C(\mathbf{s})=
\text{Pr}(y=1\mid\mathbf{s})
\in [0,1],
\label{eq:classifier_score}
\end{equation}
which we define following Refs.~\cite{scikit-learn-threshold,scikit-learn-calibration}.
Larger values of
$C(\mathbf{s})$ indicate that the syndrome is judged by the trained
classifier to be more consistent with the class-$1$ ensemble. 
Conversely, smaller $C(\mathbf{s})$ imply a higher probability $\text{Pr}(y=0\mid\mathbf{s})=1-C(\mathbf{s})$ to belong to the low-error dataset $y=0$.

The model parameters are learned by
minimizing the empirical binary cross-entropy,
\begin{equation}
\mathcal{L}(\boldsymbol{\theta})
=
-\frac{1}{N}
\sum_{j=1}^{N}
\left[
y_j \log C(\mathbf{s}_j)
+
(1-y_j)
\log\left(1-C(\mathbf{s}_j)\right)
\right],
\label{eq:binary_cross_entropy}
\end{equation}
where $N$ is the number of training samples. %
Eq.~\eqref{eq:binary_cross_entropy} represents the standard
probabilistic binary-classification formulation underlying the proposed
approach. Further details on the classifier, the class definitions, and the training and calibration procedures are provided in the
Supplementary Information, Sec.~\ref{sec:sm_ml}.

Although $C(\mathbf{s})$ is obtained as a class probability, the post-selection procedure primarily relies on the ordering of the syndromes induced by this score. Therefore, perfect probability calibration is not required, provided that syndromes associated with higher-error regimes tend to receive larger values of $C(\mathbf{s})$.

\subsection{Training data}
\label{seq:training_data}

The training data consist of two syndrome data sets,
\begin{equation}
    \mathcal{D}_{\mathrm{low}} = \{\synd_j^{\mathrm{low}}\}_{j=1}^{N_{\mathrm{low}}},
    \qquad
    \mathcal{D}_{\mathrm{high}} = \{\synd_j^{\mathrm{high}}\}_{j=1}^{N_{\mathrm{high}}},
\end{equation}
obtained at physical error rates $p_{\mathrm{low}}$ and $p_{\mathrm{high}}$, respectively.  
The values are chosen so that $p_{\mathrm{low}}$ is representative of an easy regime and $p_{\mathrm{high}}$ is a noisier regime in which logical failure is more common. 
The choice of $p_{\mathrm{low}}$ and $p_{\mathrm{high}}$ is flexible: For example, for codes with threshold, $p_{\mathrm{low}}$ is chosen below and $p_{\mathrm{high}}$ above the error correction threshold $p_\text{th}$. For experiments, one can choose it to be below and above the operating error regime of the device. 

\begin{figure*}[t!]
    \centering
    \subfigimg[width=0.95\textwidth]{}{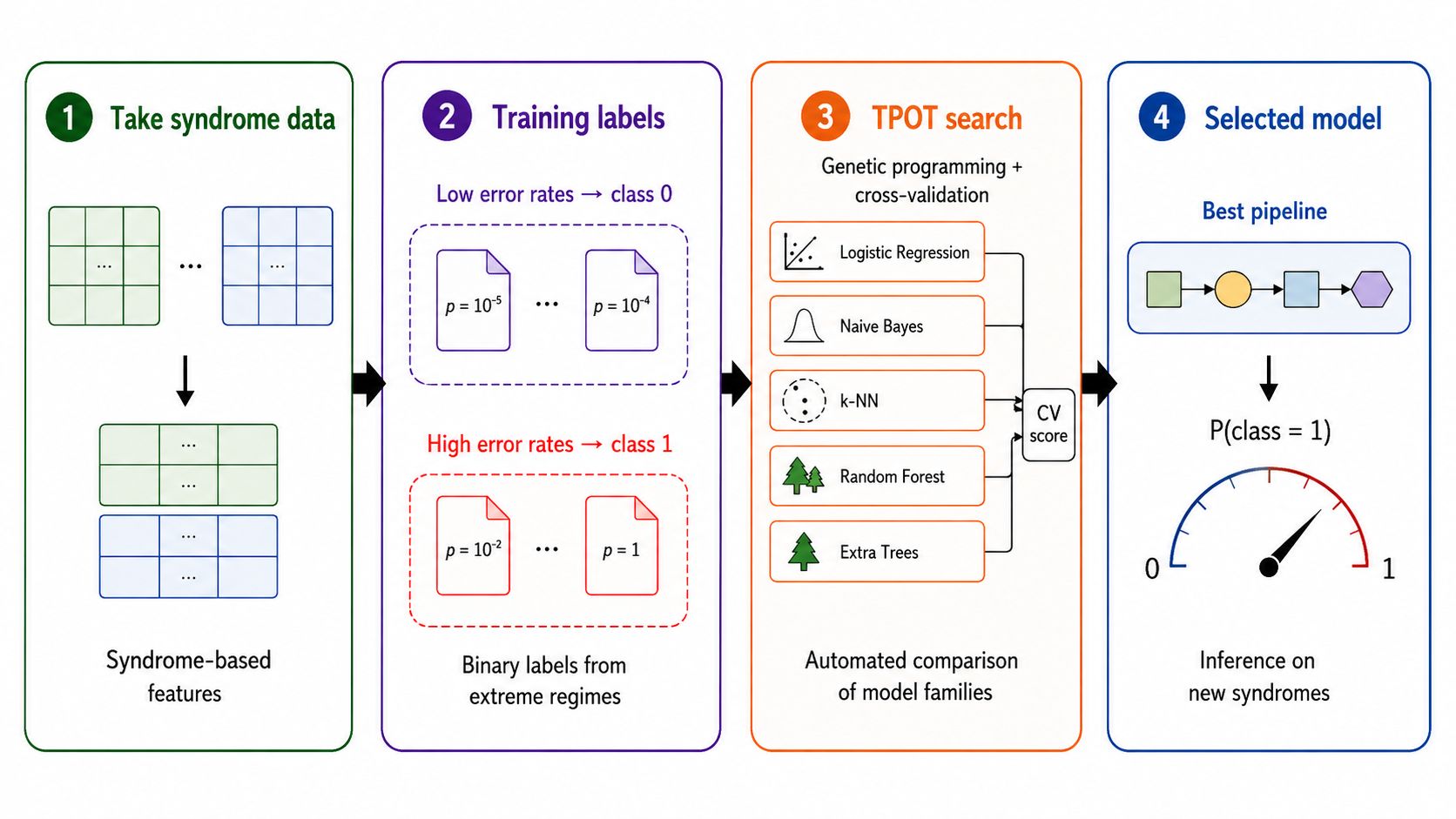}
    \caption{
{\bf Sketch of the automated machine-learning pipeline}. Syndrome records
sampled from low- and high-error-rate regimes are assigned class labels
$0$ and $1$, respectively, and used to train a binary classifier through
the TPOT automated model-selection procedure. Candidate pipelines are
evaluated by cross-validation, and the best-performing pipeline is
refitted on the training data. For each new syndrome $\mathbf{s}$, the
selected classifier returns the class-$1$ probability
$C(\mathbf{s})=\text{Pr}(y=1\mid\mathbf{s})$, which quantifies
its similarity to the high-error-rate ensemble. This score is compared
with a cutoff $C_0$, where for $C(\mathbf{s})<C_0$ we accept the run, and else reject. 
}
    \label{fig:pipeline}
\end{figure*}

Our method can be applied to both simulations and experiments of quantum error correction.
Simulated training data for realistic noise models can be efficiently generated: For stabilizer codes with Pauli noise, the quantum evolution is Clifford and thus can be efficiently simulated due to the Gottesman-Knill theorem~\cite{gottesman1998heisenberg}. 
In fact, one can efficiently sample syndrome data from the full error correction circuit with noisy gates via the tableau formalism. This can be done in time complexity $O(mn)$ where $m$ is the number of gates and $n$ the number of qubits of the code~\cite{gidney2021stim}. By varying the error probabilities of the gates, one can generate both the low and high-error-rate dataset efficiently.
For experiments, the low- and high-noise ensembles can be generated by calibrated noise amplification, by using simulations fitted to calibration data, or by combining experimental and synthetic data.  
We highlight that in contrast to other machine learning approaches~\cite{bauch2024learning,lange2025datadriven,maan2025machine,gicev2025fully,dentelski2026neuralnetworkdecoderconfidence}, our method does not require information about what logical errors have occurred, thus simulated data can be  acquired cheaply, and experimental data can be directly used for training without the need for a reference decoder.

\subsection{Model architecture}

To construct the supervised learning model to represent $C(\synd)$, we employ an automated machine-learning strategy based on Tree-based Pipeline Optimization Tool (TPOT) (see Fig.~\ref{fig:pipeline})~\cite{OlsonGECCO2016,ribeiro2024tpot2}. TPOT formulates model selection as a genetic programming problem, in which candidate machine-learning pipelines are evolved and evaluated according to a predefined performance metric. Each pipeline includes preprocessing operations, feature transformations, and a classification algorithm to explore different hyperparameter configurations, distinct model families and data-processing strategies.

We use TPOT to identify an effective binary classifier capable of discriminating between samples associated with low- and high-error-rate regimes. The input features are constructed from the syndrome information, while the target labels are assigned according to the selected low- and high-error-rate training instances. Candidate pipelines are compared using cross-validation on the training set, and the best-performing pipeline is subsequently used to assign probabilistic predictions to the remaining data.

This approach provides a systematic and reproducible framework to find the best model for the machine learning problem. 
Importantly, the TPOT search is not restricted to a single model architecture: in general, the optimization procedure can distinguish among several classes of models, including linear classifiers, tree-based ensembles, nearest-neighbor methods, and probabilistic classifiers, depending on the specified search space. Therefore, the final selected pipeline should be interpreted as the result of an automated comparison among competing hypotheses for the structure of the classifier. Detailed information about the TPOT configuration, search space, optimization parameters, and selected pipeline is provided in the \SM{}.

\subsection{Abort rule}

\begin{figure*}[ht!]
    \centering
    \subfigimg[width=0.28\textwidth]{a}{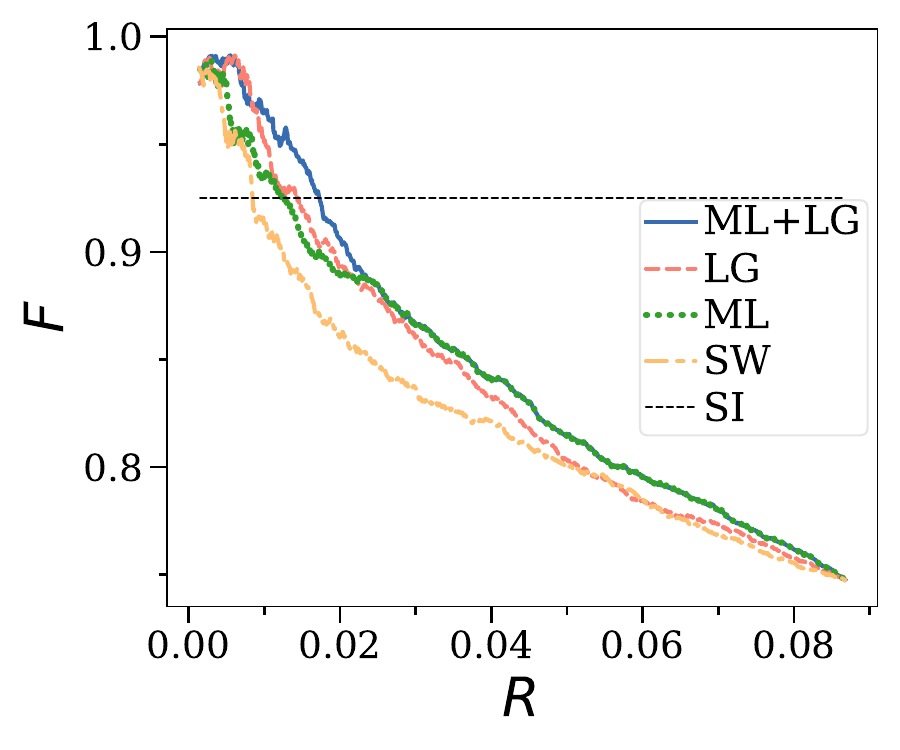}
    \subfigimg[width=0.28\textwidth]{b}{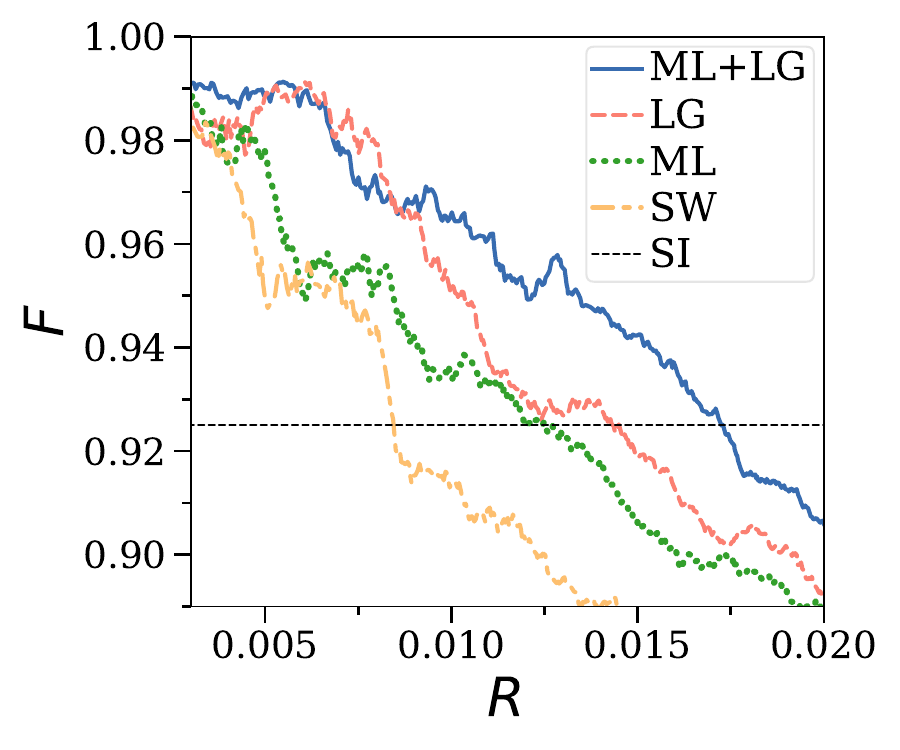}
    \subfigimg[width=0.28\textwidth]{c}{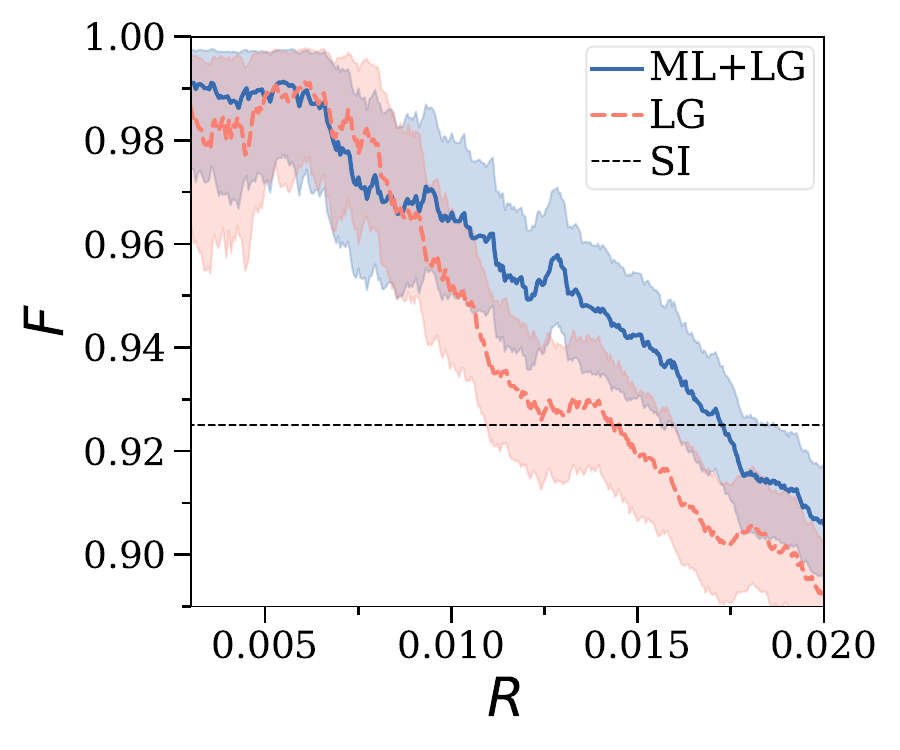}
    \caption{{\bf Post-selection enhancement for QuEra's magic-state distillation experiment}.  Experimental implementation of 5-1 magic state distillation protocol by Ref.~\cite{bravyi2005universal} where $5$ noisy single-qubit magic states are encoded into distance$-5$ color code blocks and then distilled into a less noisy magic state. The experimental protocol is described in Ref.~\cite{sales2025experimental} and the data from Ref.~\cite{quera_computing_and_collaborators_2025_15981647}.
    We show logical-error reduction by different post-selection strategies: based on the logical gap (LG), on machine-learning (ML), syndrome-weight (SW), and on machine-learning %
    combined with logical gap (ML+LG). As a reference, the black dashed line is direct state-injection (SI) into the QEC code without distillation and post-selection as reported in Ref.~\cite{sales2025experimental}.
    \idg{a} Fidelity $F$ of distilled magic state against post-selection rate $R$ for different post-selection methods.
    ML achieves higher fidelity than SW for all $R$. Compared to LG, ML performs better for larger $R$, while being slightly worse for smaller $R$. We highlight that combined ML+LG performs better than LG (as well as SW and ML).
    \idg{b} Close-up of panel a for $R\leq 0.02$. %
    \idg{c} ML+LG and LG including $68\%$ confidence intervals as shaded areas.
    The fidelity and the confidence interval are computed following the Bures prior as described in Ref.~\cite{sales2025experimental}.
    For ML, we train on simulated syndrome data which are generated via Stim, where the error rates of each gate, which were previously calibrated for the experiment, are rescaled by a factor $E$, with $E=\{0.01,0.05,0.10,0.15\}$ for low-error and $E=\{1.75,2.00,2.25,2.50\}$ for high-error ensemble, and then apply the trained model to the experimental data. After balancing the two classes, the simulated dataset contains $809{,}064$ syndromes in total, with $404{,}532$ samples in each class; we use a stratified $10\%$/$90\%$ split for training and validation, respectively. For ML+LG, we use ML to discard $80\%$ of the data, and then the remaining data is post-selected via the logical gap. }
    \label{fig:quera}
\end{figure*}

After training, the best classifier is chosen and applied to new syndrome records.  We define a cutoff $C_0\in[0,1]$, where we accept a run with syndrome $\synd$ when
\begin{equation}
    \cpred(\synd)<C_0.
    \label{eq:acceptance_rule}
\end{equation}
Thus, syndromes that look too similar to the high-noise class are rejected.  By varying $C_0$, we can map out the performance curve as function of the post-selection rate $R$, %
defined in~\eqref{eq:postselectionrate}, and the conditional logical error rate $%
\tilde{p}_L$  after the usual decoder is applied to the accepted shots.

For memory experiments we report the improvement factor
\begin{equation}
    I=\frac{\pL}{%
    \tilde{p}_L}.
    \label{eq:improvement_factor}
\end{equation}
At fixed $R$, larger $I$ means better conditional logical performance.  For magic-state data we report the conditional output fidelity $F$ as a function of $R$.

\section{Application I: Experimental Logical magic-state distillation data}
\label{sec:quera}
First, we apply our method to the recent QuEra experiment for logical magic-state distillation~\cite{sales2025experimental}.  
Here, we directly enhance the actual experiment by applying our scheme to experimental data which can be found at Ref.~\cite{quera_computing_and_collaborators_2025_15981647}.
In this experiment, the goal is not merely to preserve a memory, but to prepare a high-fidelity logical magic state via magic-state distillation. 
The experiment uses the 5-to-1 magic state distillation protocol by Ref.~\cite{bravyi2005universal}: $5$ noisy single-qubit magic states are distilled into a less noisy magic state. The scheme starts by encoding noisy magic states into  $d=5$ color codes, then the distillation protocol is applied, which includes post-selection on certain logical measurement outcomes. On successful distillation, logical tomography is used to measure the fidelity of the distilled magic state. 
As distillation requires post-selection by default, our own post-selection method can be naturally integrated to further enhance the fidelity of the produced magic states. %

The training data is generated using the Stim circuit that is calibrated on the actual experiment, where the Stim circuit can be found at Ref.~\cite{quera_computing_and_collaborators_2025_15981647}.  
Two simulated data sets are produced at physical error rates below and above the error rates calibrated to the experiment.  The machine-learning classifier is trained on these simulated syndromes and then applied to actual experimental syndrome data.
As baseline, we consider the syndrome-weight strategy, as well as computing the logical gap of most likely and second-most likely error by solving the corresponding mixed-integer problem over the decoding graph. We note that the original paper considered the logical gap for its fidelities~\cite{sales2025experimental}, where it was noted that the logical gap is quite computationally expensive to compute and scales badly with $d$. %

Fig.~\ref{fig:quera} compares four post-selection scores: logical gap (LG), machine-learning (ML), syndrome-weight (SW), and  machine-learning post-selection followed by logical gap (ML+LG).  
We also show the fidelity $F_\text{SI}\approx0.951$ for direct state-injection (SI) into the QEC code without distillation and post-selection as reported in Ref.~\cite{sales2025experimental}.

We show fidelity $F$ against post-selection rate $R$ in Fig.~\ref{fig:quera}. We find that for $R>0.025$, ML and ML+LG have the highest fidelities, even superior to LG alone. 
In Fig.~\ref{fig:quera}b, we show the close-up for low $R$. We find that for small $R$, ML+LG and LG perform similarly well. Interestingly, even ML alone is superior to SW, thus showing that our machine learning model learns information beyond syndrome-weight from the data. 
This allows ML to perform quite well, while having much lower complexity than LG. In particular, ML alone already exceeds the fidelities achieved by direct injection of the magic state without magic-state distillation, thus demonstrating that sufficient performance can be achieved without high-cost logical-gap calculation.
Finally, Fig.~\ref{fig:quera}c shows ML+LG and LG with $68\%$ confidence intervals, showing that our results are robust. We highlight that ML+LG is superior to direct state injection for $R^{\text{ML+LG}}\approx0.017$, which is an improvement over the result of LG only at $R^{\text{LG}}\approx0.014$~\cite{sales2025experimental}.

\section{Application II: Gross bivariate-bicycle code}
\label{sec:gross}

Next, we study the $[\![144,12,12]\!]$ Gross bivariate-bicycle code, a qLDPC code with weight-6 checks~\cite{bravyi2024high}.  The simulations %
use a circuit-level depolarizing-noise model following Ref.~\cite{bravyi2024high}.  This setting is useful because it probes the method beyond the standard surface-code decoding graph and includes measurement and circuit faults, not only data-qubit errors.

The classifier is trained on syndrome data sampled from high and low physical error rates.   
Then, the cutoff $C_0$ in Eq.~\eqref{eq:acceptance_rule} is swept to obtain different post-selection rates $R$.  The accepted shots are then decoded using the same decoder as in the unfiltered baseline.

\begin{figure}[ht!]
    \centering
        \subfigimg[width=0.24\textwidth]{a}{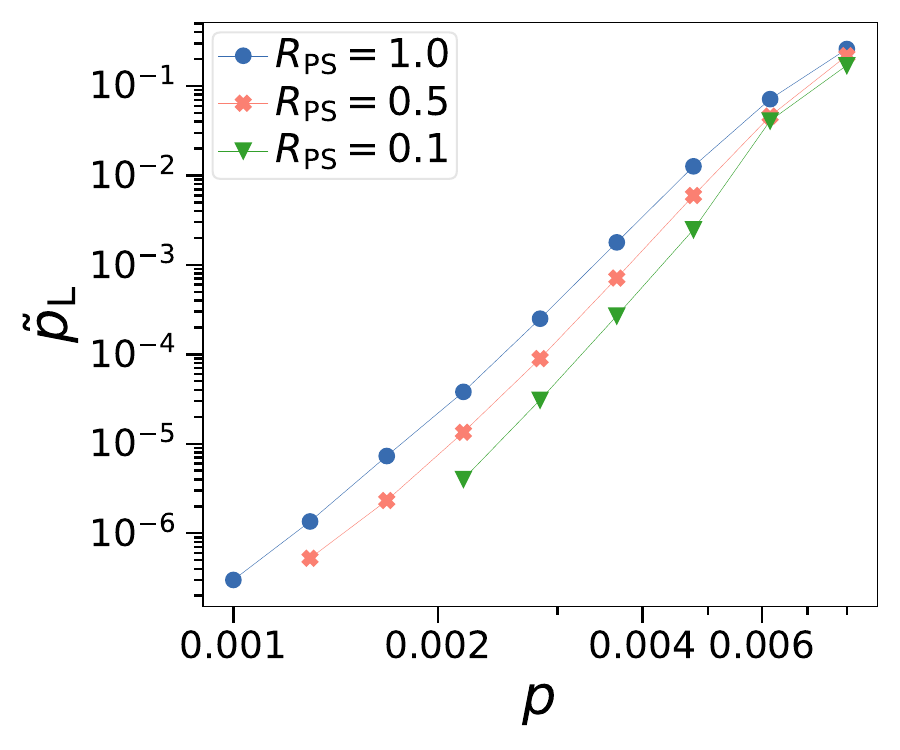}\hfill
    \subfigimg[width=0.24\textwidth]{b}{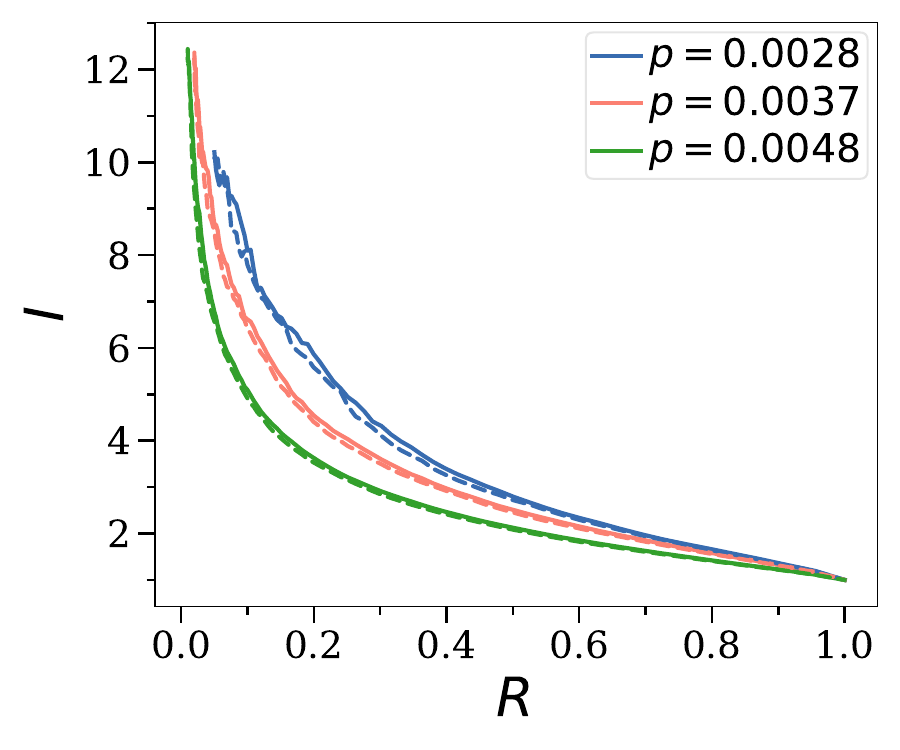}
    \caption{{\bf Syndrome post-selection enhancement of the %
    Gross-code}. 
    For the $[\![144,12,12]\!]$ bivariate-bicycle code, we train a classifier model on an efficiently generated dataset of syndromes drawn at physical error rates $p=\{0.001 , 0.0013\}$ (low-noise) or $p=\{0.0062, 0.008\}$ (high-noise). 
    We use the trained model to post-select syndromes at each $p$ and then decode with BP-OSD.
    \idg{a} Logical error rate $%
    \tilde{p}_L$ after %
    post-selection  at post-selection 
    rate $R$ (defined in~\eqref{eq:postselectionrate})  %
    as a function of $p$. As expected, the lower $R$ is, the lower $\tilde{p}_L$ is. %
    \idg{b} Improvement factor $I=\pL/%
    \tilde{p}_L$ plotted against $R$ for different $p$. 
    Solid lines correspond to our %
    post-selection via machine learning%
    , while dashed lines to %
    the syndrome-weight criterion. 
    The training set was %
    a stratified $15\%$ subset of the $2{,}199{,}720$ available syndrome records, corresponding to $329{,}958$ training samples ($283{,}104$ from the low-error regime and $46{,}854$ from the high-error one) with $2{,}016$ features each. 
    The logical rates reported correspond to average rates per syndrome-extraction cycle, obtained from the total rates over $12$ cycles as $\pL = 1-(1-\pL^{(12%
    )})^{1/12}$ and $\tilde{p}_L = 1-(1-\tilde{p}_L^{(12%
    )})^{1/12}$ where $\pL^{12}$ is the logical error rate after $12$ cycles.
    }
    \label{fig:gross}
\end{figure}

In Fig.~\ref{fig:gross}, we show the improvement of machine-learning post-selection for the Gross code. In Fig.~\ref{fig:gross}a, we show $%
\tilde{p}_L$ obtained via machine learning against $p$, finding that the logical error rate improves with decreasing $R$ relative to the non-postselection baseline at $R=1$.
In Fig.~\ref{fig:gross}b, we show improvement $I$ over the baseline against $R$ for different $p$. We show machine learning post-selection as solid line, while dashed line is the syndrome-weight method. 
For the Gross-code, the machine-learning and syndrome-weight strategies show broadly similar trends for simple decoding graphs. This highlights that the machine learning model learns a similar strategy as counting flipped syndromes, which seems to already perform quite well.

\section{Application III: Surface-code post-selection transition}
\label{sec:surface}

Finally, we consider the rotated surface code $[\![d^2,1,d]\!]$, which provides a controlled setting in which to study scaling with code distance.  We consider a code-capacity model where every qubit is subject to bit-flip noise with probability $p$ and measurements are assumed to be noise-free. %

\begin{figure}[ht!]
    \centering
    \subfigimg[width=0.24\textwidth]{a}{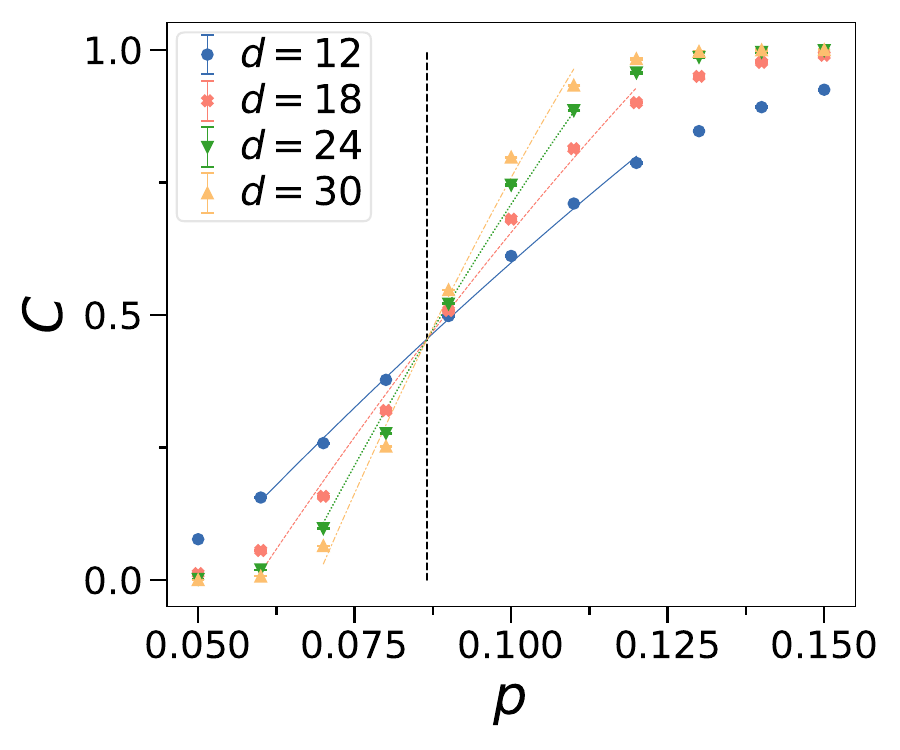}\hfill
    \subfigimg[width=0.24\textwidth]{b}{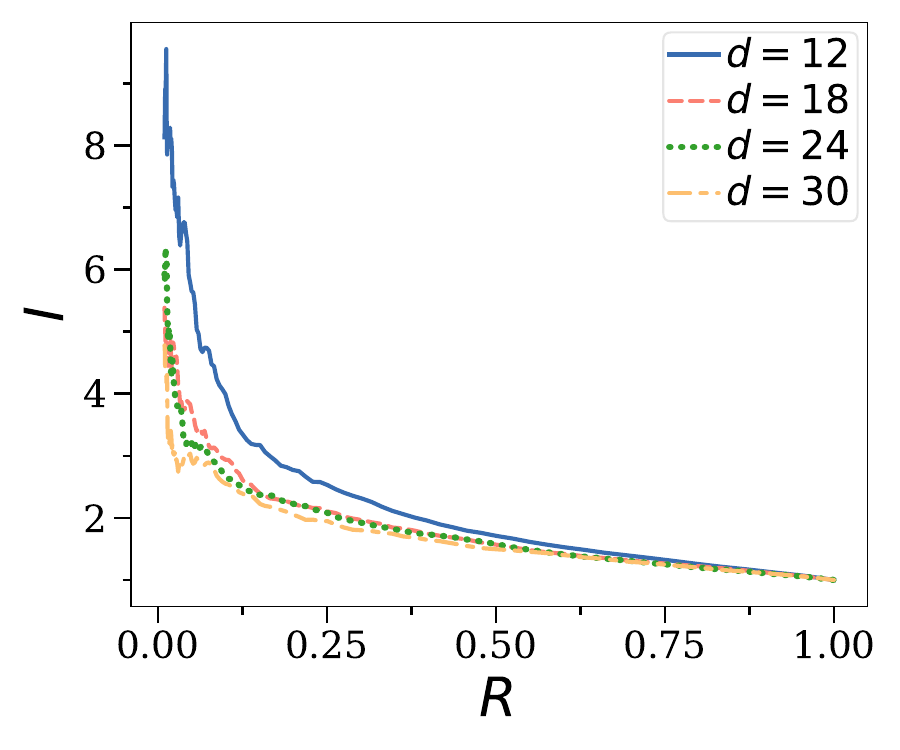}
    \caption{{\bf Surface-code classifier threshold}.  
    \idg{a} The machine-learning predictor $\cpred$ against $p$, for different code distances $d$. 
    For all $d$, the curves cross at $\pml\approx 0.0866(1)$ (vertical dashed line) and exhibit universal behavior (see \SM{}~\ref{sec:collapse}), reminiscent of a phase transition at a post-selection probability threshold. 
    Interestingly, $\pml$ differs from the usual physical-error threshold $\pth$ of the code (see main text). 
    For the model's training, we used syndrome data generated at physical error rates $p=0.05$ and $p=0.06$, for low noise, and $p=0.14$ and $p=0.15$, for high noise.
    Approximately $10^{5}$ syndrome samples are available at each physical error rate, where we use $20\%$ of the data for training and $80\%$ for validation, while the reported predictor values are averaged over $10^{4}$ syndrome samples per error rate. We train the classifier for each $d$ individually.
    \idg{b} Improvement $I$ with post-selection against post-selection rate $R$ for $p\approx\pml$ and different values of $d$. We decode with %
    MWPM.
    }
    \label{fig:surface_threshold}
\end{figure}

We find that the trained machine-learning classifier $C$ that distinguishes low and high-error ensembles exhibits a finite-size crossing consistent with a phase transition. In particular, in Fig.~\ref{fig:surface_threshold}a we plot $C$ against $p$ for different $d$. The curves of $C$ for different $d$ cross near the point $\pml\approx 0.0866$. In fact, after finite-size rescaling, the data approximately collapse for all $d$ onto a single curve, highlighting the universal behavior expected from a phase transition (see \SM{}~\ref{sec:collapse}). 
This implies that for large $d$, syndromes generated below this point are nearly always classified as low-noise-like, while syndromes generated above this point are classified as high-noise-like and are preferentially rejected.  
Notably, the observed crossing is distinct from the logical-error threshold $\pth\approx 0.1037$, where we find $\pml<\pth$ which we show in \SM{}~\ref{sec:collapse}.

This behavior suggests that machine-learning post-selection can detect a data-driven transition in the syndrome distribution.  The transition is not identical to the decoding threshold: it is a property of distinguishability between two syndrome ensembles as represented by the learned classifier.  
Such a transition has also been observed for syndrome-weight post-selection method proposed in Ref.~\cite{english2025thresholds}.

The transition identifies a scalable post-selection method that works even for large $d$: At $p\approx \pml$,  sweeping the cutoff $C_0$ produces a scalable post-selection curve whose improvement factor $I=\pL/%
\tilde{p}_L$ is nearly independent of distance $d$.  
We confirm this in Fig.~\ref{fig:surface_threshold}b, where we show $I$ against post-selection rate $R$ for different $d$ at the post-selection transition. We show both machine learning and syndrome-weight method, which show similar improvement.
We find that for $d\geq18$, the relationship between $I$ and $R$ is nearly independent of $d$. Thus, we expect that we can reliably improve the logical error rate with post-selection by a constant factor as $d$ increases.

\section{Discussion}
\label{sec:discussion}

We introduced a machine-learning method based on syndrome data for post-selecting quantum error correction runs.  Our method trains a machine-learning classifier on syndrome data from low- and high-noise ensembles and uses the classifier output as a pre-decoding abort score.  Because training does not require decoded logical labels or information about whether a logical error has occurred, the method avoids one of the main obstacles to machine-learning-assisted QEC.  

Our method can serve as a low-cost pre-decoding filter. For each syndrome record, the classifier first estimates how strongly the record resembles the high-noise training ensemble. Runs assigned a sufficiently high score can be rejected without invoking the full decoder, while accepted runs are passed to the standard decoding pipeline. This can reduce classical decoding workload and improve the conditional logical performance of the accepted ensemble, at the cost of a reduced acceptance probability. Such a trade-off is particularly relevant to probabilistic protocols, including magic-state distillation~\cite{bravyi2005universal,sales2025experimental}, magic-state cultivation~\cite{gidney2024magic}, and repeat-until-success gadgets~\cite{akahoshi2024partially}, in which high conditional fidelity may be more important than deterministic output. In the experimental magic-state-distillation data, the learned score outperforms syndrome-weight filtering, indicating that it captures structure beyond the total number of detection events.

As baseline, we compare our method against the syndrome-weight post-selection method~\cite{li2015magic,singh2022high,bombin2024fault,english2025thresholds}, which aborts when the number of flipped syndromes exceeds a threshold. This method performs comparable to our machine-learning method for the Gross code and surface code, indicating that data-driven methods learn a classifier that behaves similarly to counting the number of syndromes.
As the surface code and Gross code simulations use relatively simple bit-flip and circuit-level depolarising noise models, respectively, syndrome counting already contains sufficient information for post-selection. 
However, for the magic-state distillation experiment, our machine learning post-selection method outperforms simple syndrome counting, implying that by learning from data, we can identify more reliably which runs should be aborted. 
As the magic-state distillation experiment uses a complex and realistic noise model with highly non-trivial relationships between syndromes and errors,  counting syndrome events does not reliable indicate logical failures. 
In contrast, our machine learning model can learn the structure in the syndrome distribution to more reliably decide when to post-select.

We also observe a post-selection transition $\pml$ distinct from the usual logical threshold $\pth$ for the surface code. 
In fact, the machine-learning predictor $C$ exhibits universal behavior, evident by a universal collapse of the $C$ against $p$ curve under proper rescaling for different $d$, as shown in \SM{}~\ref{sec:collapse}. Around $p\approx\pml$, our post-selection method is scalable as the post-selection probability remains non-trivial even for large $d$.
A similar post-selection transition has been observed for the syndrome-weight post-selection method~\cite{english2025thresholds}.
Thus, we show that machine learning can autonomously learn this post-selection transition.

Our method has several practical advantages.  First, it is modular: it can be inserted before any existing decoder without changing the decoding pipeline.  In fact, we find that by first performing machine learning post-selection followed by logical gap post-selection, we can achieve superior fidelity compared to logical gap alone for a recent magic state distillation experiment by QuEra~\cite{sales2025experimental}.

Second, it is flexible: the same principle applies to memory experiments, qLDPC codes, topological codes, magic state distillation, and logical state-preparation experiments.  
In fact, our decoder can be trained on both simulated and experimental data, as long as the physical error can be tuned to generate the high and low-error dataset. 

Finally, our machine learning approach shows transferability from simulation to experiment, illustrated by the magic-state distillation application where we train on simulated data, then apply to experimental data.  We believe it would be interesting to apply our approach to other recent experiments~\cite{rosenfeld2025magic,gupta2024encoding,bluvstein2024logical,google2025quantum}.

\begin{acknowledgements}
    We thank Andr\'e J. Ferreira-Martins for discussions. We are very grateful to Tommaso Macrì and Chen Zhao from QuEra for extensive and helpful discussions.  %
    AC also thanks UFAL for an unpaid license and CNPq (Grant No. 168785/2023-4).
\end{acknowledgements}

\bibliography{refs}

\let\addcontentsline\oldaddcontentsline

\appendix

\onecolumngrid

\newpage 

\setcounter{secnumdepth}{2}

\renewcommand{\thesection}{\Alph{section}}
\renewcommand{\thesubsection}{\arabic{subsection}}

\clearpage
\begin{center}

\textbf{\large Supplemental Information}
\end{center}

In the Supplemental Material, we provide additional analyses and methodological details.

\tableofcontents

\section{Post-selection and error correction threshold}\label{sec:collapse}
\begin{figure*}[htbp]
    \centering
    \subfigimg[width=0.3\textwidth]{}{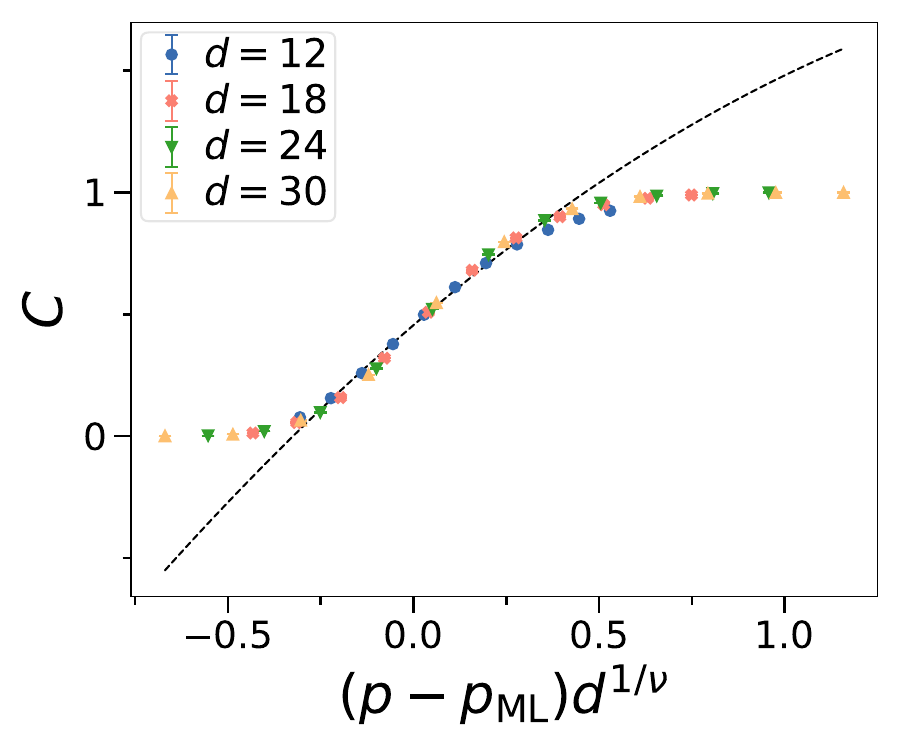}
    \caption{Surface-code classifier threshold.  
     The machine-learning predictor $\cpred$ against $p$ with fitted selection probability threshold  $\pml\approx 0.0866(1)$.  By rescaling, curves for different $d$ can be collapsed onto a single curve around the post-selection probability threshold $\pml$, demonstrating the universality of $\cpred$. The dashed line is the fitted function~\eqref{eq:fit}.
    }
    \label{fig:surface_threshold_collapse}
\end{figure*}

In this section, we show for the surface code the explicit collapse of machine-learning predictor $C$  onto a single curve by appropriate rescaling. 
To get $\pml$, we fit the data with the polynomial~\cite{watson2014logical}
\begin{equation}\label{eq:fit}
    C=A+Bx+Dx^2 \quad\text{with}\quad x = (p-\pml)d^{1/\nu}.
\end{equation} 
For the fitting, we remove data with very small and large $p$ which are too far from the threshold $p_\text{ML}$.
Then, we rescale in Fig.~\ref{fig:surface_threshold_collapse} the physical error rate $p$ as $(p-\pml)d^{1/\nu}$, observing that close to the threshold indeed all curves have universal behavior, i.e. are a simple function of $d$. 

\begin{figure*}[htbp]
    \centering
    \subfigimg[width=0.3\textwidth]{b}{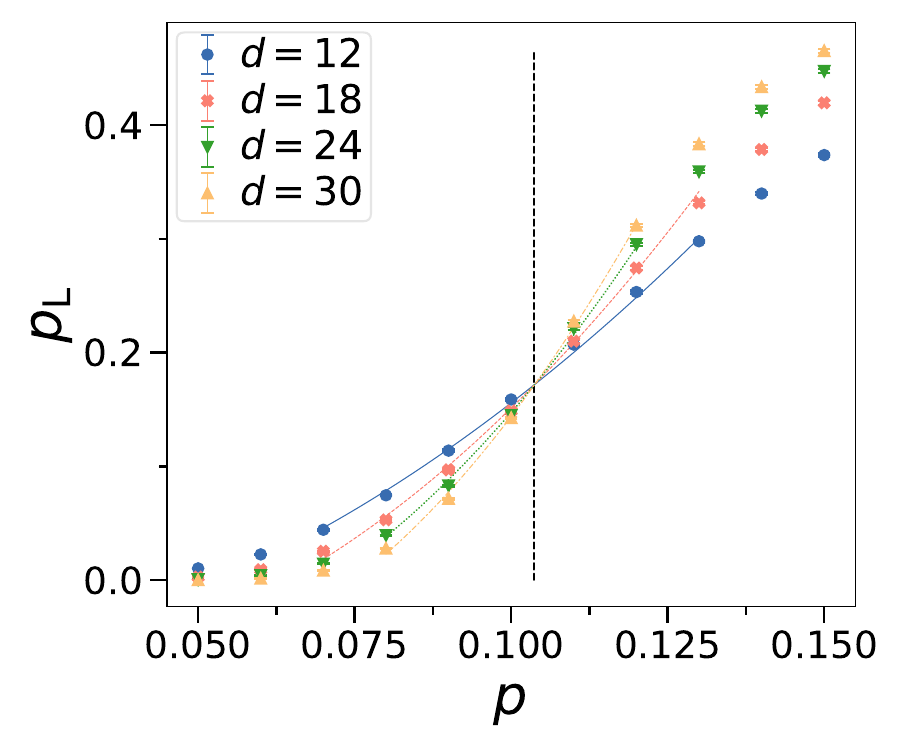}
    \caption{As reference, we show the (non-post-selected) logical error rate $\pL$ against physical error rate $p$ for different distances $d$, with fitted threshold $\pth\approx 0.1037(2)$ marked as vertical dashed line.  
    }
    \label{fig:surface_threshold_sup}
\end{figure*}

We also note that the post-selection threshold is distinct from the usual error correction threshold. As reference, we show the usual error correction threshold by fitting logical error $\pL$ against physical error $p$ in Fig.~\ref{fig:surface_threshold_sup}, where we find $\pth\approx 0.1037(2)$ where indeed we have $\pth>\pml$.

\section{Machine-learning methodology}
\label{sec:sm_ml}

In this section, we provide further details on the construction,
training and application of the machine-learning model used for
syndrome post-selection.

\subsection{Dataset construction}
As an example, we describe the dataset generation for our Gross bivariate-bicycle code application.
For the Gross bivariate-bicycle code, each input sample was constructed
from the syndrome information associated with a single simulated QEC
run. The syndrome arrays corresponding to the two error components were
concatenated into a single feature vector. Each syndrome record was
therefore represented by $2016$ binary input features.

The target labels identify the physical-error-rate ensemble from which
each syndrome was generated. The two lowest physical error rates,

\begin{equation}
p \in
\left\{
0.001,\,
0.0013
\right\},
\end{equation}

were assigned to class $0$, corresponding to the low-error-rate
ensemble. The two highest physical error rates,

\begin{equation}
p \in
\left\{
0.0062,\,
0.008
\right\},
\end{equation}

were assigned to class $1$, corresponding to the high-error-rate
ensemble.

The complete labeled dataset contained $2{,}199{,}720$ syndrome
records. Of these, $1{,}887{,}360$ samples belonged to the
low-error-rate class and $312{,}360$ samples belonged to the
high-error-rate class.

A stratified train--test split was used to preserve the relative class
frequencies. We used $15\%$ of the complete dataset for model training
and retained the remaining $85\%$ for evaluation. The training subset
therefore contained $329{,}958$ syndrome records, comprising
$283{,}104$ low-error-rate samples and $46{,}854$ high-error-rate
samples. The split was performed using random seed $7$.

\subsection{Automated pipeline search}

The classifier was selected using the Tree-based Pipeline Optimization
Tool, TPOT, version $0.12.1$~\cite{OlsonGECCO2016,ribeiro2024tpot2}. TPOT formulates model selection and
hyperparameter optimization as a genetic-programming problem in which
candidate machine-learning pipelines are generated, evaluated through
cross-validation, and iteratively evolved according to their predictive
performance.

Our post-selection procedure requires a continuous score representing
the inferred probability that a syndrome belongs to the high-error-rate
class. We therefore restricted the TPOT search space to classification
models to learn the predicted probability $C(\mathbf{s})=
\text{Pr}(y=1\mid\mathbf{s})$. This restriction
ensures that every candidate pipeline returns class probabilities and
can consequently provide the class-$1$ score.

The restricted search space included logistic regression, random
forests, gradient-boosting classifiers, Gaussian naive Bayes,
$k$-nearest-neighbor classifiers, and support-vector classifiers with
probability estimation enabled. The hyperparameter search space was
defined as follows:

\begin{itemize}

\item Logistic regression with regularization terms
$\Lambda \in \{10^{-4},10^{-3},10^{-2},10^{-1},1,10,100\}$,
$\ell_2$ regularization, and the \texttt{lbfgs} solver.

\item Random forests with
$100$ or $200$ estimators,
maximum depth in $\{\mathrm{None},5,10\}$,
minimum split size in $\{2,5\}$,
and minimum leaf size in $\{1,2\}$.

\item Gradient-boosting classifiers with
$100$ or $200$ estimators,
learning rate in $\{0.01,0.1\}$,
and maximum tree depth in $\{3,5\}$.

\item Gaussian naive Bayes using its standard hyperparameters.

\item $k$-nearest-neighbor classifiers with
$k \in \{3,5,7\}$.

\item Support-vector classifiers with
$\Lambda \in \{0.1,1,10\}$,
either a linear or radial-basis-function kernel,
and probability estimation enabled.

\end{itemize}

The TPOT optimization was performed using a population size of $10$
over $5$ generations. The TPOT random seed was set to $11$. No explicit
scoring function or cross-validation scheme was provided, and therefore
the defaults of TPOT version $0.12.1$ were used.

The best internal cross-validation score reached $1.0$. The pipeline
selected by TPOT was an $\ell_2$-regularized logistic-regression
classifier with
\begin{equation}
\Lambda=1,
\end{equation}
using the \texttt{lbfgs} solver. After model selection, the selected
pipeline was refitted on the full training subset and frozen before
being applied to the remaining syndrome data.

\subsection{Probabilistic inference}

For each new syndrome record $\mathbf{s}$, the selected classifier
returns the estimated probabilities of the two training classes. We define the machine-learning score
as the inferred probability of class $1$,
\begin{equation}
C(\mathbf{s})
=
\text{Pr}(y=1\mid\mathbf{s}),
\end{equation}

where class $1$ denotes the high-error-rate ensemble. Larger values of
$C(\mathbf{s})$ indicate that the observed syndrome is more
characteristic of the high-error-rate training data, whereas smaller
values indicate greater similarity to the low-error-rate ensemble.

The model is not trained to predict whether decoding succeeds, whether
a logical error occurs, or which correction operator should be applied.
Instead, the class-$1$ probability provides a scalar ranking of syndrome
records according to their similarity to the high-error-rate ensemble.

\subsection{Post-selection rule}

For a chosen cutoff $C_{0}\in[0,1]$, a QEC run with syndrome
$\mathbf{s}$ is accepted when
\begin{equation}
C(\mathbf{s}) < C_{0},
\end{equation}
and rejected otherwise. Syndromes assigned a sufficiently large
probability of belonging to class $1$ are therefore discarded before
decoding.

By varying $C_{0}$, we control the trade-off between the fraction $R$ of
accepted runs and the conditional logical error rate $\tilde{p}_\text{L}$ of the retained
ensemble. The classifier is frozen during this analysis, and only the
post-selection cutoff is varied.

Although the score is returned as a class probability, the
post-selection procedure primarily depends on the ordering induced by
$C(\mathbf{s})$. 
Thus, perfect calibration of the score is not required for ranking syndromes, instead we only need that syndromes characteristic of higher-error
regimes tend to receive larger class-$1$ scores.

\end{document}